\documentclass[twocolumn,noshowpacs,preprintnumbers,amsmath,amssymb]{revtex4}

\usepackage{verbatim}
\usepackage{wasysym}

\usepackage{graphicx}
\usepackage{dcolumn}
\usepackage{bm}
\usepackage{wrapfig}

\begin{document}

\title{Reversible attachment of platinum alloy nanoparticles to non-functionalized carbon nanotubes}

\author{Beate Ritz}
\author{Hauke Heller}
\author{Anton Myalitsin}
\author{Andreas Kornowski}
\affiliation{Institute of Physical Chemistry, University of Hamburg, 20146 Hamburg, Germany}
\author{Francisco J. Martin-Martinez}
\author{Santiago Melchor}
\author{Jose A. Dobado}
\affiliation{Department of Organic Chemistry, University of Granada, 18071 Granada, Spain}
\author{Beatriz H. Juarez}
\affiliation{IMDEA Nanoscience, 28049 Madrid, Spain}
\author{Horst Weller}
\author{Christian Klinke}
\email{klinke@chemie.uni-hamburg.de}
\affiliation{Institute of Physical Chemistry, University of Hamburg, 20146 Hamburg, Germany}

\begin{abstract} 

The formation of monodisperse, tunable sized, alloyed nanoparticles of Ni, Co, or Fe with Pt and pure Pt nanoparticles attached to carbon nanotubes has been investigated. Following homogeneous nucleation, nanoparticles attach directly to non-functionalized singlewall and multiwall carbon nanotubes during nanoparticle synthesis as a function of ligand nature and the nanoparticle work function. These ligands do not only provide a way to tune the chemical composition, size and shape of the nanoparticles but also control a strong reversible interaction with carbon nanotubes and permit controlling the nanoparticle coverage. Raman spectroscopy reveals that the sp2 hybridization of the carbon lattice is not modified by the attachment. In order to better understand the interaction between the directly attached nanoparticles and the non-functionalized carbon nanotubes we employed first-principles calculations on model systems of small Pt clusters and both zig-zag and armchair singlewall carbon nanotubes. The detailed comprehension of such systems is of major importance since they find applications in catalysis and energy storage.

\end{abstract}

\maketitle

Composites of metallic nanoparticles (NPs) and carbon nanotubes (CNTs) exhibit high catalytic activity for various chemical reactions \cite{1,2,3,4,5,6} and have also been explored for hydrogen storage applications\cite{7}. Recent reports include platinum \cite{6,8,9}, cobalt \cite{10,11}, nickel \cite{7,12}, gold \cite{13}, rhodium \cite{14} as well as platinum nickel \cite{15}, platinum tin \cite{16} and platinum ruthenium \cite{17} NPs immobilized on CNTs \cite{18}. Alloying NPs of platinum with nickel, cobalt, or iron allows tunable magnetic response \cite{19,20} and catalytic properties \cite{21,22,23}. Beyond that, 1D alignment of NPs enables to modify the saturation magnetization and coercitivity through magnetostatic coupling \cite{24,25,26}. A convenient way for 1D alignment is the attachment of NPs to CNTs, which is usually achieved by electrochemical deposition \cite{27,28}, the reduction of metallic salts in the presence of functionalized CNTs \cite{15,29}, or chemical vapor deposition \cite{10}, among others \cite{30}. On the other hand, concerning the NP synthesis, the organometallic synthesis route provides nanocrystalline alloyed materials with precise size control and tunable composition in several systems \cite{20,31,32}. Here, we report on the synthesis of alloyed $Ni_{x}Pt_{1-x}$ \cite{20}, $Co_{x}Pt_{1-x}$ and $Fe_{x}Pt_{1-x}$ \cite{32} NPs as well as pure $Pt$ \cite{33} NPs and their attachment to non-functionalized singlewall (SWCNTs), multiwall carbon nanotubes (MWCNTs) and glassy carbon by their simple integration in the organometallic synthesis. The experimental procedure involves only a single synthetic step, whereby the crucial parameter for attachment was found in the correct balance of the ligands oleylamine (OA) and oleic acid (Oac). Under the appropriate conditions, only poorly stabilized NPs are formed which enables the CNTs to act as additional ligands as it was previously observed for the system CdSe-CNTs \cite{34,35,36}. Additionally, we present first-principle calculations indicating a charge transfer from the CNTs to the NPs upon attachment resulting in positive charging of the CNTs. These findings are understood in terms of Fermi-level equilibration leading to an electrostatic stabilization of the composite material.

The composites were synthesized by the reduction or thermal decomposition of the metal precursors in the presence of CNTs or glassy carbon, OA or a mixture of OA and Oac, and 1,2-hexadecanediol in diphenyl ether at 200$^{\circ}$C (see experimental section for details). We have synthesized a large variety of NPs with different sizes and compositions. Some examples are listed in Table I.

\begin{table*}[tbp]
\begin{center}
\caption {\textit{Optimized ligand to metal ratios (in 10 mL diphenyl ether in the presence of 2 mg CNTs) for different metal precursors and NPs composition.}}
\vspace{0.5cm}
\begin{tabular}{|c|c|c|c|c|c|c|c|c|} \hline
\# & Me precursor & Oac [mmol] & Oac/Me & OA [mmol] & OA/Me & OA/total Me (incl. Pt) & $\diameter$ [nm] & Composition \\ \hline

1 & $Ni(ac)_{2}$ & - & - & 0.15 & 1 & 0.5 & 5.5 $\pm$ 0.7 & $Ni_{46}Pt_{54}$ \\ \hline

2 & $Co_{2}(CO)_{8}$ & 6.3 & 38 & 0.15 & 1 & 0.5 & 4.9 $\pm$ 0.9 & $Co_{9}Pt_{91}$ \\ \hline

3 & $CoCl_{2} \cdot 6H_{2}O$ & - & - & 0.15 & 1 & 0.5 & 6.9 $\pm$ 1.2 & $Co_{19}Pt_{81}$ \\ \hline

4 & $CoCl_{2} \cdot 6H_{2}O$ & - & - & 0.30 & 2 & 1 & 9.3 $\pm$ 1.7 & $Co_{31}Pt_{69}$ \\ \hline

5 & $Fe(CO)_{5}$ & - & - & 0.15 & 1 & 0.5 & 3.4 $\pm$ 0.4 & $Fe_{40}Pt_{60}$ \\ \hline

6 & $Pt(acac)_{2}$ & 0.15 & 1 & 0.15 & 1 & 1 & 2-10 & $Pt$ \\ \hline

\end{tabular}
\end{center}
\end{table*}

All alloy samples consist of monodisperse NPs covering densely the surface of the CNTs. Such high coverage was also observed for pure Pt NPs, which are, however, rather polydisperse. To follow the process in more detail we took aliquots from the reaction mixture after different times. Representative TEM images are shown in Fig. 1 for the NiPt-CNT system after 1 and 10 minutes of reaction. As reported for the particle synthesis in homogeneous solution the particle formation is fast and the reaction is usually finished within a few minutes after the injection of the platinum precursor \cite{20,32,38,39,40}. Accordingly, we find a large amount of small particles either freely in solution or covering the CNTs already after 1 minute of synthesis. In the next minutes the particles grow both, in solution and on the CNTs, which are homogeneously and densely covered with NPs after 10 minutes (Fig. 1b). It is, therefore, most likely that homogeneous nucleation occurs in solution followed by an attachment to the CNTs, although partial nucleation on the CNTs cannot be excluded. Such composites are stable at least over months and endure ultrasound treatment for 24 h.

\begin{figure}[htbp]
  \centering
  \includegraphics[width=0.45\textwidth]{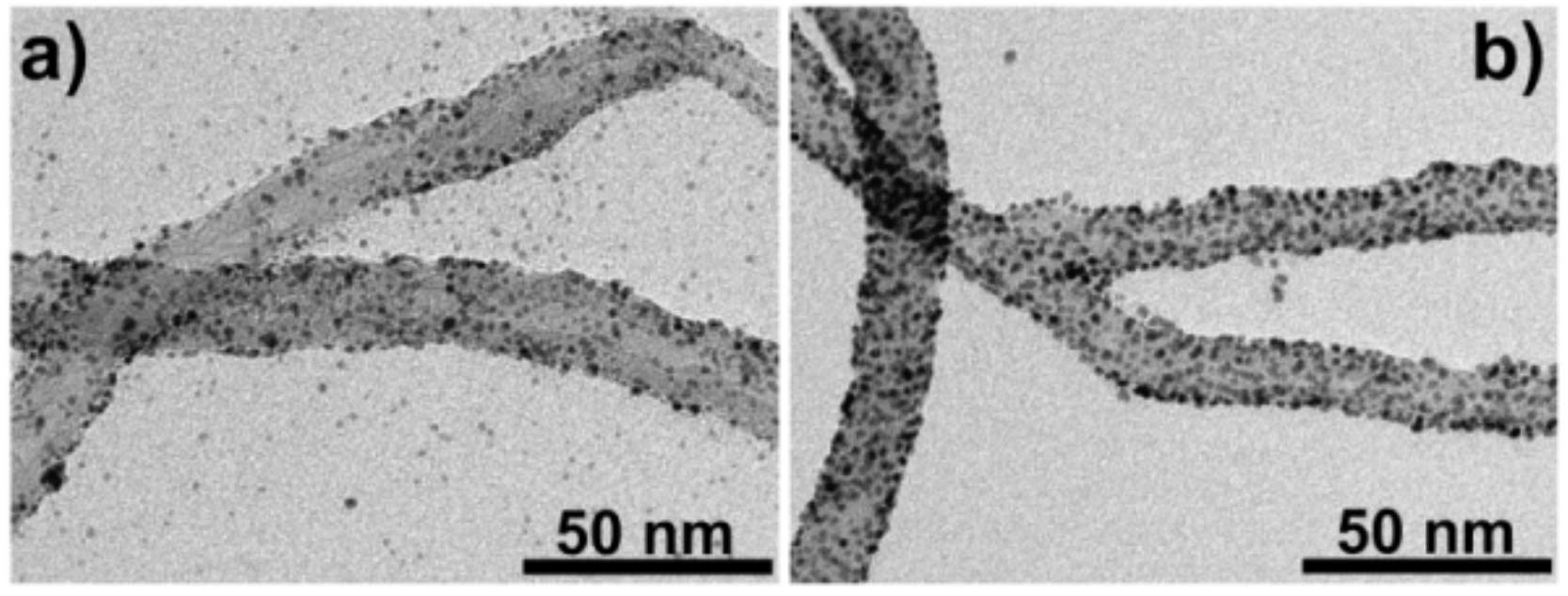}
  \caption{\textit{TEM images of NiPt-CNT composites 1 minute (a) and 10 minutes (b) after the injection of the platinum precursor.}}
\end{figure}

HRTEM images of CoPt NPs attached to SWCNTs (Fig. 2) show an atomically close proximity of the NP and the CNT surface while there is a visible distance of about 0.8 nm between two adjacent NPs due to the ligands on the NP surfaces. The NPs are closely connected to the CNTs surface with no apparent ligands in between. Thus, the facet connected to the CNTs may lack ligands allowing direct connection between the NP surface and the CNT. Analysis of the crystal orientation reveals that the particles attach via the (100) facet. This seems plausible when one keeps in mind the nearly cubic shape of the particles. Detailed investigations on small, almost spherical particles show, however, that the attachment also occurs selectively via the (100) facet. 

\begin{figure}[htbp]
  \centering
  \includegraphics[width=0.45\textwidth]{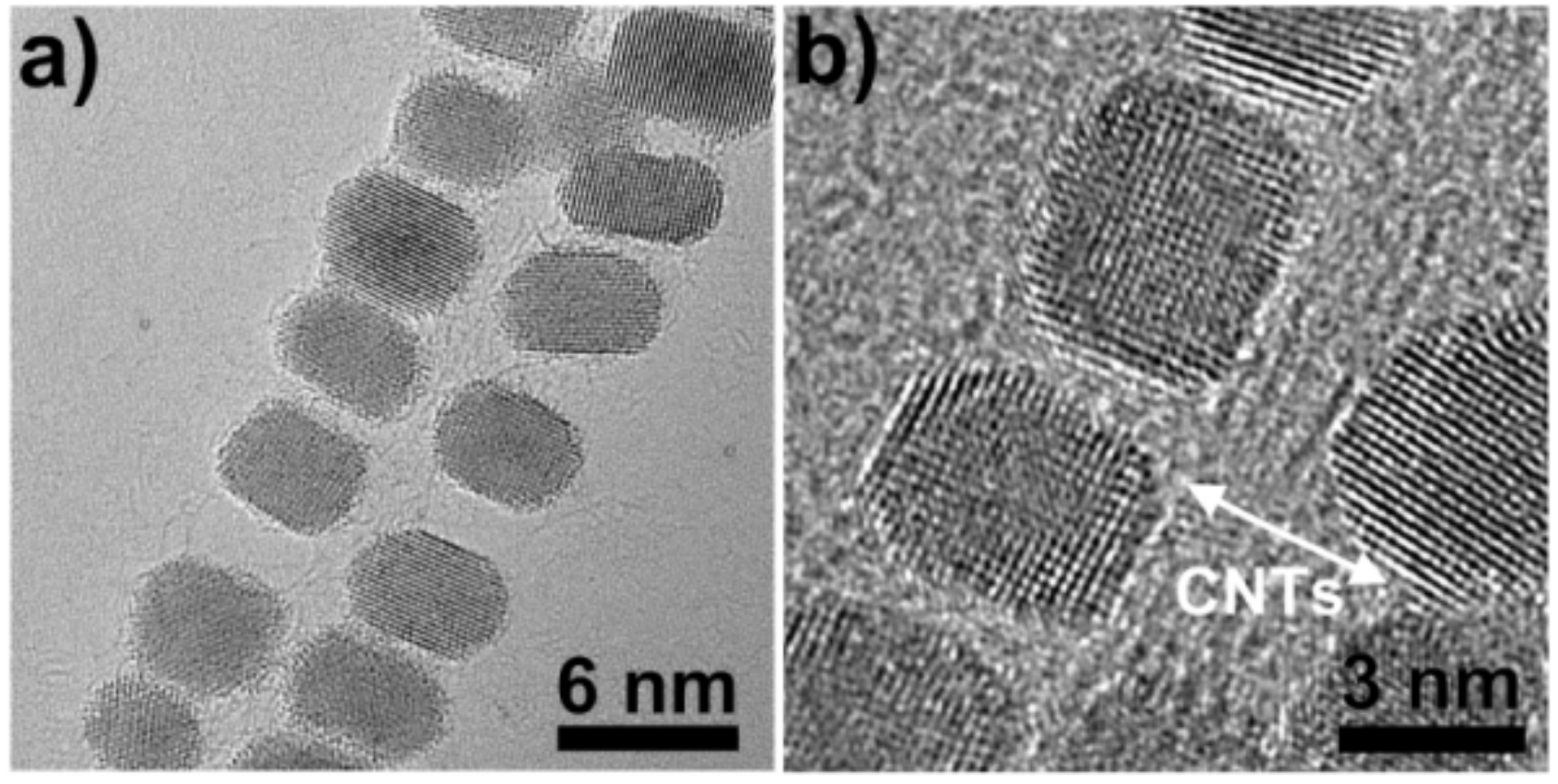}
  \caption{\textit{HRTEM images of CoPt NPs attached to SWCNTs.}}
\end{figure}

The role of Oac and OA on the nucleation and growth of platinum alloy particles has been investigated by several authors \cite{20,38,40,41,42,43}. Consistently, it was found that OA is essential to obtain stable colloidal solutions of nanoparticles showing defined size, shape and composition, whereas Oac influences their composition and size. The non-existing contribution of Oac on the attachment that we found in our experiments allows producing highly covered NP-CNT composites with tunable size and composition of the NPs. Further on, we found that OA additionally controls the immobilization of the NPs on the CNTs. This can be clearly seen in the TEM images in Fig. 3. While comparably low OA concentrations (see Experimental Section for details) yield highly covered CNTs even if a high Oac concentrations is used (Fig. 3a), higher concentrations of OA produce NP aggregated or separated in solution (Fig. 3b) preventing further attachment.

\begin{figure}[htbp]
  \centering
  \includegraphics[width=0.45\textwidth]{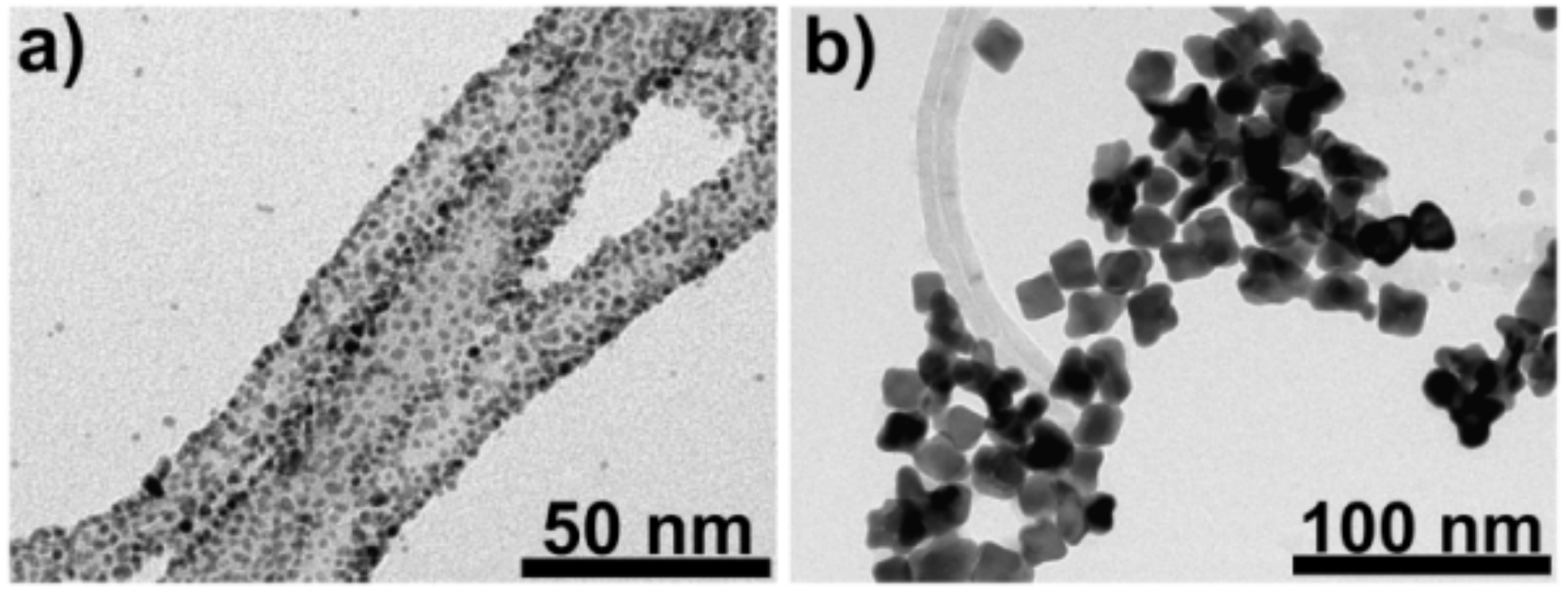}
  \caption{\textit{NiPt based NP-CNT composites synthesized with (a) 6.3 mmol Oac and 0.15 mmol OA and (b) 6.1 mmol OA.}}
\end{figure}

It is worth mentioning that, the addition of an excess of OA to a sample with high coverage of NPs (prepared with 0.15 mmol OA) results in spontaneous detachment of the particles from the CNTs (Fig. 4b). In contract, the addition of the same amount of Oac does not influence the attachment significantly (Fig. 4a). Accordingly, it is not only possible to remove the NPs from the surface of CNTs but also to attach them in a post-synthesis procedure. NPs insufficiently stabilized by OA mixed with CNTs yield similar composites due to the stabilization of the NPs by the CNTs (see Experimental Methods).

\begin{figure}[htbp]
  \centering
  \includegraphics[width=0.45\textwidth]{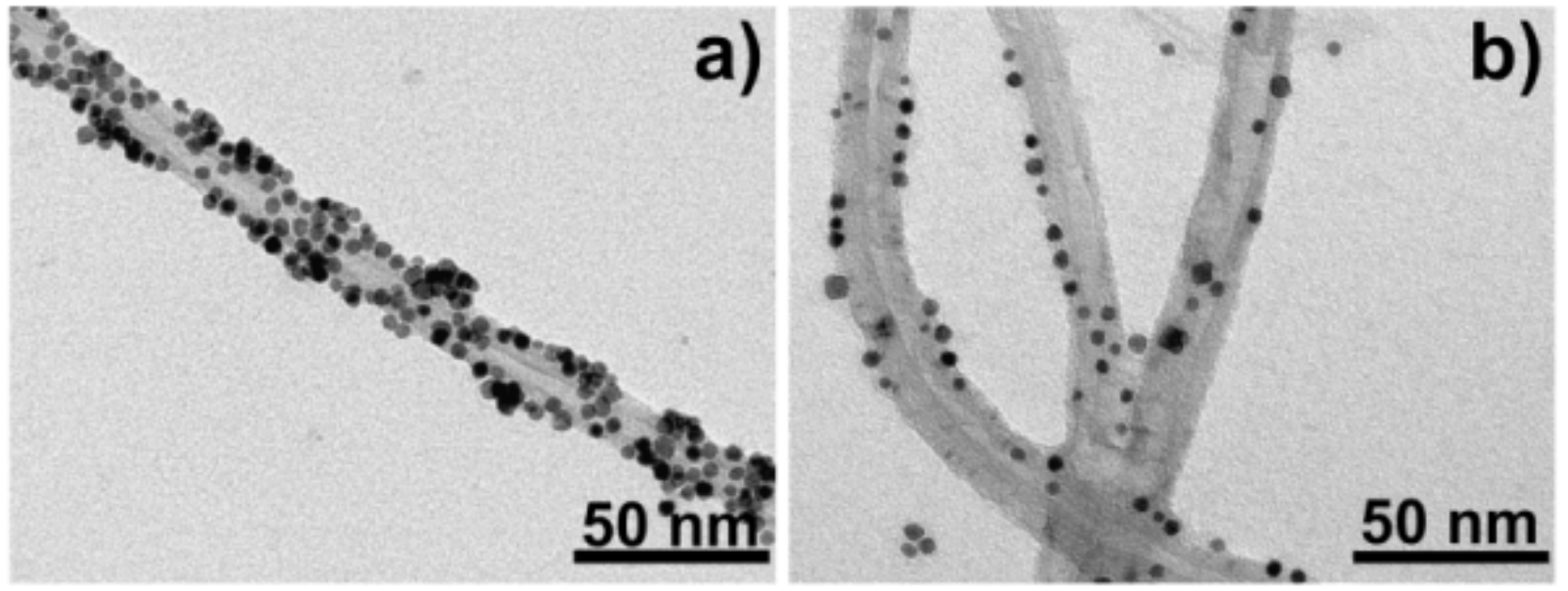}
  \caption{\textit{NiPt MWCNT composites (Fig. 1b) sonicated with Oac (a) and same amount of OA (b).}}
\end{figure}

We therefore conclude that the strong binding of OA to Pt competes with the NPs anchoring to CNTs. In this sense NP attachment to the graphene-like system CNTs can be understood as a ligand exchange process as already described for a CdSe-CNT system \cite{35}. 

In order to learn more about the type of interaction between NPs and CNTs Raman spectroscopy was performed. The here analyzed features are the disorder-induced mode (D-band) and the graphene like tangential displacement mode (G-band), as shown in Fig. 5 for the Nanocyl SWCNTs used in this study. SWCNTs were probed using an Ar-Kr laser at two different wavelengths at 514.5 nm (2.41 eV) and 647.1 nm (1.92 eV). The lower wavelength was primarily resonant with the semiconducting fraction of the Nanocyl SWCNTs employed in this study while the higher wavelength was mainly resonant with the metallic fraction. The D mode corresponds to the sp$^{3}$ hybridization of the CNTs and is sensitive to covalent functionalization \cite{44}. The observed intensity changes of the D-band relative to the sum of the D- and G-band upon immobilization of NPs were within the error bars of the respective experiments, both in the cases of primarily probing semiconducting and metallic SWCNTs. If covalent functionalization of the CNT structure occurred, the intensity of the D-band would increase significantly. We therefore consider these changes negligible and conclude that mainly non covalent bonding took place. Thus, the sp$^{2}$ structure of the CNTs is preserved during the formation of NP-CNT composites.

\begin{figure}[htbp]
  \centering
  \includegraphics[width=0.45\textwidth]{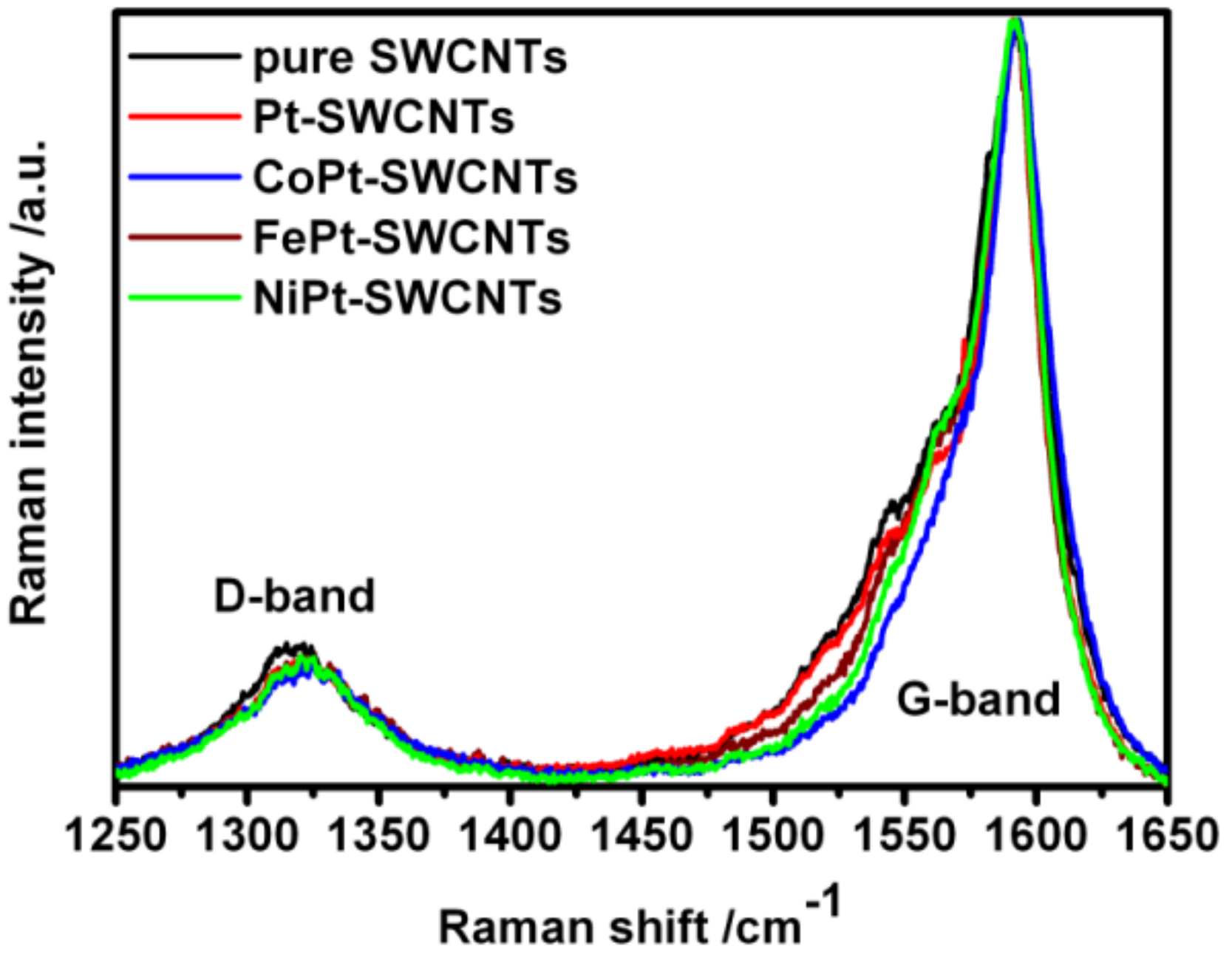}
  \caption{\textit{D and G band of pristine SWCNTs and NP-SWCNT composites with high loads of FePt, CoPt, NiPt and Pt NPs using a laser energy of 2.41 eV.}}
\end{figure}

Further information about the nature of the interaction between NPs and CNTs can be gained by first-principles theoretical calculations. We modeled systems of small Pt clusters on both zigzag and armchair CNTs. The interaction between Pt and CNTs has been previously analyzed in the literature employing tight-binding \cite{45,46} and plane-wave calculations \cite{47}. However, for a detailed analysis advanced tools for bond nature characterization such as Quantum Theory of Atoms in Molecules (QTAIM) \cite{48} or Electron Localization Function (ELF) \cite{49} are necessary. Thus, we calculated optimized geometries and total energies by Density Functional Theory (DFT) methods, employing the hybrid B3LYP functional \cite{50,51}, together with the LanL2DZ ECP basis set with effective core potentials for Pt atoms \cite{52,53,54,55,56} and the STO-3G one for C and H atoms \cite{57,58}. In order to analyze the charge transfer and the bonding nature of the interaction Molecular Electrostatic Potential (MEP), QTAIM, and ELF49 have been employed. Due to a more accurate description of the electron density of localized electrons in bonds, we chose a finite molecular model rather than a periodic system. As a matter of fact, finite and periodic models differ sometimes in the resulting geometry of isolated CNTs \cite{59}. Thus, we checked the current geometric results with the reported periodic ones \cite{47}. The final model system consists of a pyramidal cluster composed by 14 Pt atoms over the surface of a (10,0) CNT (see Figure 6) \cite{60}. We have also explored other CNTs such as (12,0), (5,5) and (6,6) for comparison, but the results are similar in all cases. Therefore, we focus the discussion on (10,0) CNTs. 

First, we fully optimized all the different configurations between one Pt atom and the (10,0) CNT to obtain the preferred absorption spots and the Pt-C equilibrium distance, which has been compared with the results using periodic boundary conditions. One Pt atom can be placed over a C atom of the CNT, above a bond, or above a center of a carbon ring. However, due to the CNT topology, two different bonds have to be considered, those in the axis direction and those in a certain angle with the axis direction, resulting in four different configurations. Optimization showed that the Pt atoms above C-C bonds are the preferred position with a Pt-C distance of 2.0 $\AA$  in accordance with Ref. \cite{47}. Subsequently, we constructed a 3x3 atom sheet in $\left\{100\right\}$ and $\left\{111\right\}$ fcc facets configuration over a CNT with an angle of 0$^{\circ}$, 30$^{\circ}$, and 45$^{\circ}$ relative to each other. We explored the Potential Energy Surface with a single point energy calculations scan for 24 different combinations \cite{61} using the Pt-C distance obtained in the previous step. Using this nine Pt atom cluster (1 shell) in $\left\{100\right\}$ and $\left\{111\right\}$ fcc arrangement, the calculations showed that the interaction with the $\left\{100\right\}$ facets are the preferred ones, supporting the experimental results. The lowest energy situation is the one in which the central atom in the nine atom $\left\{100\right\}$ facets is placed on top of the bond in axis direction and the angle between the cluster and the CNT is 45$^{\circ}$. This situation allows a high number of Pt atoms to match on top of bonds and C atoms (preferred positions) rather than on rings. The minimum energy configuration was further optimized with a fixed 14 Pt atoms cluster pyramid (using 2.7 $\AA$ for the Pt-Pt distance). With this model system, it is observed that the CNT is slightly deformed towards the Pt cluster. This provides evidence for the attraction between Pt and CNTs. Moreover, this deformation is highlighted not only from a geometrical point of view but also from an electronic one. In Figure 6a, the MEP plot shows clearly more negative values around the platinum cluster in interaction with the CNT (red isosurface), which is even more noteworthy if we look at the MEP distribution of the isolated cluster (inset of Figure 6a), in which a non homogeneous distribution is presented. This fact, together with the emptied MEP isosurface in the region of the CNT opposite to the cluster (in yellow), suggests a charge transfer from the CNT to the cluster, as we expected since the value of the Pt work function is higher than the one of C. 

\begin{figure}[htbp]
  \centering
  \includegraphics[width=0.45\textwidth]{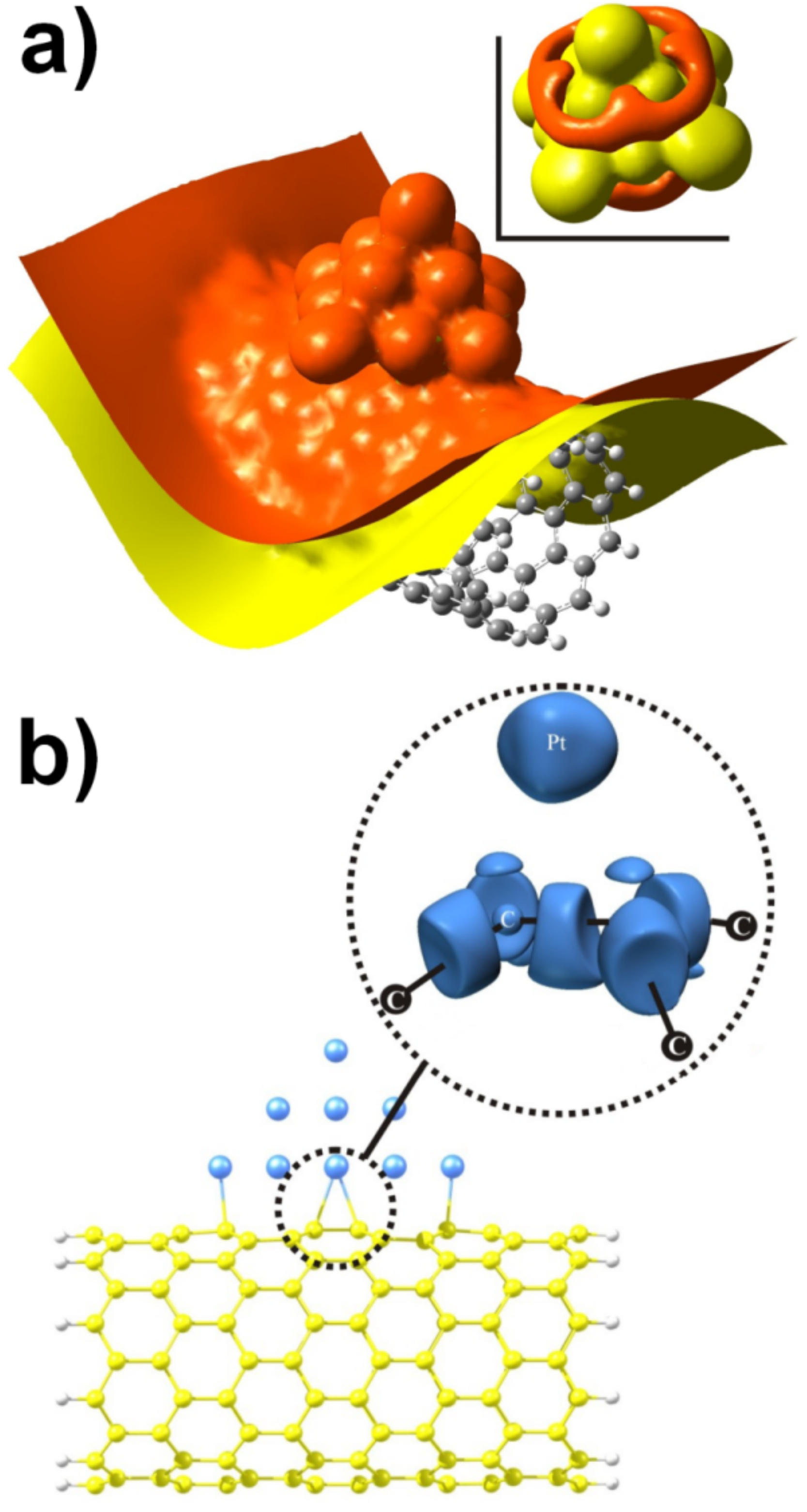}
  \caption{\textit{Models of 14 Pt atoms on a (10,0) SWCNT. a) MEP plot of the composite and an isolated Pt cluster as inset. Red indicates a surface of constant negative potential and yellow a positive one. b) With an ELF analysis showing the paired electron density in blue. Small kidney-shaped islands can be found between the carbon lattice and the Pt cluster, indicating a minor covalent contribution. In contrast strong covalent bonds are localized in between the carbon atoms.}}
\end{figure}

The work functions of semiconducting and metallic SWCNTs are 4.8-5.4 eV and 4.5-5.0 eV, respectively \cite{62}. While Pt features a work function of 5.65 eV (averaged values from Landolt-Bornstein) those of Ni and Co are considerably lower with 5.15 and 5.0 eV. Finally, the value of Fe is even lower with 4.5 eV. The resulting work functions for the alloys should be in between the value of platinum and the respective metal, depending on the composition and crystal structure of the alloys. The computational results together with the values for the work functions show that charge transfer from the SWCNTs to the platinum alloys is plausible unless the iron content is too high. Calculated binding energies for single atoms to CNTs further support our results. Spin polarized binding energies of 2.4 eV for a single platinum atom, 1.7 eV for a single cobalt or nickel atom and finally 0.8 eV in the case of a single iron atom are reported63. On the other hand, ELF calculations show weak electron localization in the path from C to Pt (Fig. 6b), which implies a small covalent contribution to the bond, although too small and sparse to be seen in Raman spectroscopy. Thus, according to theoretical calculations, the $\left\{100\right\}$ fcc facet is the most stable one for the interaction, arising from a charge transfer from the CNT to the metal that rules the attachment with a small covalent contribution, causing a geometrical deformation of the CNT structure. 

Summarizing, we have shown that fine tuning of the concentration of oleic acid and oleylamine, allows control not only over the size, shape, and composition of Pt-based nanoparticles but also on their reversible attachment to CNTs. Computational investigations provided evidence for a charge stabilization of the composites. The work functions of platinum based NPs are lower than those of the CNTs leading to a charge transfer from the CNTs to the NPs until the Fermi levels are equilibrated. The different polarization of the materials provides the attractive forces responsible for the charge stabilization, while the electronic structure of the CNTs is preserved. Such composites may find applications in fuel cells or energy storage. The knowledge about the involved mechanism of reversible attachment might help to develop composite systems with NPs of different nature.

\

\textbf{ACKNOWLEDGEMENTS}

\

This work was financially supported by the "`Landesexzellenzinitiative Spintronics"' (Hamburg), the Deutsche Forschungsgemeinschaft (GK 661), the Spanish "`Ministerio de Ciencia e Innovacion"'  (CTQ2007-65112, MAT2009-13488), European Commission (ERG-2008-239256), and the "`Ramon y Cajal"' programme. We also thank the "`Centro de Servicios de Informatica y Redes de Comunicaciones"' (CSIRC), University of Granada, for the use of the UGRGrid computing facilities.  

\

\textbf{EXPERIMENTAL DETAILS}

\

The synthesis was carried out under a nitrogen atmosphere. In a typical synthesis of spherical Ni40Pt60 particles attached to CNTs $Ni(ac)_{2}$ (41.6 mg), 1,2-hexadecanediol (43.0 mg), oleylamine (0.05 mL) and diphenyl ether (8 mL) were mixed in a three-neck flask equipped with a reflux condenser, septum, and a heat controller. The mixture was kept at 80$^{\circ}$C for one hour under vacuum conditions to remove traces of water. Then the CNTs or the glassy carbon spheres (from 1-10 mg) suspended in diphenyl ether (1-5 mL) by sonication were added and conditioned again. The mixture was then heated to 200$^{\circ}$C under nitrogen atmosphere and $Pt(acac)_{2}$ (65.5 mg) dissolved in 1,2-dichlorobenzene (0.6 mL) was injected amid vigorous stirring. The black dispersion was stirred for 2 hours. The reaction mixture was cooled to room temperature and chloroform (10 mL) was added under ambient conditions. The CNT composites were isolated by centrifugation. The precipitate was dispersed in chloroform and washed two times. 

The synthesis of $Co_{19}Pt_{81}$ CNT composites with $CoCl_{2} \cdot 6H_{2}O$ and $Fe_{40}Pt_{60}$ with $Fe(CO)_{5}$ as source of cobalt or iron is comparable. If $Co_{2}(CO)_{8}$ is used we need higher amounts of Oac to get CNT composites as described above, and the $Co_{2}(CO)_{8}$ has to be injected in the hot solution as well. 

\

\textit{Post synthesis procedure}

For the attachment of NPs in a post synthesis step, the NPs were washed after the synthesis for six times with chloroform and methanol. Then the particles were suspended in 2 mL diphenyl ether and injected in a solution containing 2 mg CNTs in 8 mL diphenyl ether and 0.05 mL oleylamine at 200$^{\circ}$C. After stirring this mixture for 2 h at 200$^{\circ}$C the composites were washed with chloroform. 

\

\textit{Ligand exchange}

Treatment of the NP-CNT composites by ultrasonication in the presence of OA leads to nearly complete detachment of the NPs already after a few minutes. If the NP-CNT composites are ultrasonicated with the same amount of Oac the NPs persist on the CNTs. Although, there is also some detachment visible, it is a lot less pronounced (TEM images shown in the Fig. 4). Without supplemental ligands the composites are stable over more than 24 h treating by ultrasonication. 

\

\textit{Methods}

The composites were characterized by high-resolution transmission electron microscopy with a JEOL JEM 1011 at an acceleration voltage of 100 KV, Philips CM 300 UT at an acceleration voltage of 200 KV, and a JEOL JEM 2200 FS (UHR) with CESCOR and CETCOR corrector at an acceleration voltage of 200 kV. The composition of the particles was determined by energy dispersive X-ray spectroscopy (EDX) with an EDAX detector (DX4) spectrometer connected to the Philips CM 300 UT. Samples for TEM analysis were prepared by drying a dispersion of the CNT composite in chloroform on a carbon coated copper grid. Sample preparation for XRD measurements involved dropping the dispersion on a single crystal Si support and evaporating the solvent. The x-ray diffractogramms were recorded on a Philips X´Pert diffractometer with Bragg-Brentano geometry and copper K-alpha radiation.

For the Raman spectroscopy measurements the composites were dropped on a silicon wafer and the chloroform evaporated. The experiments were performed using Ar-Kr laser lines of 514.5 nm (2.41 eV) or 647.1 nm (1.92 eV with a high numerical aperture Zeiss Epiplan Apochromat objective (150x, NA = 0.9). The laser power was 2.5 mW corresponding to 2 MW/cm$^{2}$ while the grating employed had 1200 groves/mm. 

\

\textit{Computational Methods}

The DFT calculations were carried out with Gaussian03 \cite{64} and NWChem 5.1 \cite{65}, while the construction of CNTs was performed with the CoNTub 1.0 program \cite{66,67}.


\begin{thebibliography}{00}
\bibitem{58} Although C atoms could be better described with a higher basis set, the minimal basis set STO-3G constitutes a suitable compromise between accuracy and computational performance, as we have previously reported that it reproduces accurately the geometries of CNTs, compared with other split-valence basis sets. See Ref. 59.
\bibitem{60} In the configuration obtained in a previous single point energy scan, w. f. o. a. P. c., with a Pt-Pt distance of 2.7 $\AA$, which had been calculated before from an optimization of the isolated cluster. In the cluster-CNT optimization we froze the coordinates of the platinum atoms, leaving the CNT free to relax. Thus, we avoided eventual deformations in the cluster and emulated a more real situation in which 14 Pt atoms are surrounded for others atoms in the NP, which restrict somehow their movement.
\bibitem{61} To make this scan systematic we took the central atom of the 9 Pt array as a reference, and place it on the four different positions already explored for one Pt atom-CNT system. The three different orientations explored are 0°, 30° and 45° angle between the cluster and CNT axis.
\end{thebibliography}
\end{document}